\documentclass[]{spie}  

\usepackage{amsmath}
\usepackage{amsfonts,amssymb}
\usepackage{graphicx}
\usepackage[acronym, nonumberlist]{glossaries}
\usepackage{multicol}
\usepackage[colorlinks=true, allcolors=blue]{hyperref}
\usepackage{caption}
\usepackage{hyperref}
\hypersetup{
    colorlinks = false, 
    pdfborder = {0 0 0}, 
}

\DeclareCaptionLabelFormat{bold}{\textbf{#1 #2}}
\newcommand{\estimates}{\mathrel{\hat{=}}}
\newacronym{pet}{PET}{positron emission tomography}
\newacronym{fdg}{[$^{18}$F]-FDG}{[$^{18}$F]Fluorodeoxyglucose}
\newglossaryentry{suvr}{
  name={SUVR},
  description={standardized uptake value ratio},
  text={SUVR},
  first={SUVR}
}
\newacronym{adni}{ADNI}{Alzheimer's Disease Neuroimaging Initiative}
\newacronym{aal}{AAL}{automated anatomical labeling}
\newacronym{roi}{ROI}{region-of-interest}

\newacronym{add}{ADD}{AD dementia}
\newacronym{mci}{MCI}{mild cognitive impairment}
\newacronym{cn}{CN}{cognitively normal}
\newacronym{cdr}{CDR-SOB}{Clinical Dementia Rating Scale Sum of Boxes}
\newacronym{mmse}{MMSE}{Mini Mental State Examination}
\newacronym{adas}{ADAS-13}{Alzheimer’s Disease Assessment Scale–Cognitive Subscale-13}

\newacronym{ad}{AD}{Alzheimer's disease}

\newacronym{cnn}{CNN}{convolutional neural network}
\newacronym{lstm}{LSTM}{long-short term memory}
\newacronym{mae}{MAE}{mean absolute error}
\newglossaryentry{ssim}{
  name={SSIM},
  description={structural similarity index},
  text={SSIM},
  first={SSIM}
}
\newacronym{relu}{ReLU}{rectified linear unit}
\newacronym{cv}{CV}{cross-validation}
\newacronym{anova}{ANOVA}{analysis of variance}
\newacronym{i2i}{I2I}{image-to-image translation}

\makeglossaries

\title{Translating the future: Image-to-image translation for the prediction of future brain metabolism}
\author[a,b]{Elena Doering}
\author[b,c]{Merle C. H\"onig}
\author[d,e]{Tobias Deu{\ss}er}
\author[b]{Gérard N. Bischof}
\author[b,f]{Thilo van Eimeren}
\author[a,b,c]{Alexander Drzezga}
\author[g]{Lotta M. Ellingsen}
\author[ ]{for the Alzheimer's Disease Neuroimaging Initiative*}

\affil[a]{German Center for Neurodegenerative Diseases (DZNE), Bonn, Germany}
\affil[b]{University of Cologne, Faculty of Medicine and University Hospital Cologne, Dept. of Nuclear Medicine, Cologne, Germany}
\affil[c]{Research Center J\"ulich, Molecular Organization of the Brain (INM-2), J\"ulich, Germany}
\affil[d]{Fraunhofer IAIS, Sankt Augustin, Germany}
\affil[e]{University of Bonn, Bonn, Germany}
\affil[f]{University of Cologne, Faculty of Medicine and University Hospital Cologne, Dept. of Neurology, Cologne, Germany}
\affil[g]{University of Iceland, Dept. of Electrical and Computer Engineering, Reykjavik, Iceland}

\authorinfo{Further author information: (Send correspondence to Elena Doering)\\Elena Doering: Email: elena.doering@uk-koeln.de}
\begin{document} 
\maketitle

\begin{abstract}
Alzheimer's disease (AD) is a progressive neurodegenerative disorder leading to cognitive decline. [$^{18}$F]-Fluoro- deoxyglucose positron emission tomography ([$^{18}$F]-FDG PET) is used to monitor brain metabolism, aiding in the diagnosis and assessment of AD over time. However, the feasibility of multi-time point [$^{18}$F]-FDG PET scans for diagnosis is limited due to radiation exposure, cost, and patient burden. To address this, we have developed a predictive image-to-image translation (I2I) model to forecast future [$^{18}$F]-FDG PET scans using baseline and year-one data. The proposed model employs a convolutional neural network architecture with long-short term memory and was trained on [$^{18}$F]-FDG PET data from 161 individuals from the Alzheimer’s Disease Neuroimaging Initiative. Our I2I network showed high accuracy in predicting year-two [$^{18}$F]-FDG PET scans, with a mean absolute error of 0.031 and a structural similarity index of 0.961. Furthermore, the model successfully predicted PET scans up to seven years post-baseline. Notably, the predicted [$^{18}$F]-FDG PET signal in an AD-susceptible meta-region was highly accurate for individuals with mild cognitive impairment across years. In contrast, a linear model was sufficient for predicting brain metabolism in cognitively normal and dementia subjects. In conclusion, both the I2I network and the linear model could offer valuable prognostic insights, guiding early intervention strategies to preemptively address anticipated declines in brain metabolism and potentially to monitor treatment effects.
\end{abstract}


\section{INTRODUCTION}
\label{sec:intro}
\Gls{ad} is biologically characterized by a range of molecular changes, beginning with abnormal $\beta$-amyloid and tau accumulation and culminating in neurodegeneration, which, in turn, leads to progressive cognitive decline and ultimately dementia. The progression of \gls{ad} from asymptomatic to dementia is marked by significant inter-individual variability, both in the rate and the spatial patterns of brain metabolic decline \cite{Caminiti2023, Mosconi2010}. This information plays a vital role in the differential diagnosis \cite{Nestor2018} and monitoring of \gls{ad} \cite{Beheshti2022}. Brain \gls{pet} with \gls{fdg} is a powerful imaging modality that provides insights into brain metabolism by measuring neuronal dysfunction. Monitoring an individual's \gls{fdg} \gls{pet} scans over the time course of several years allows to track changes in brain function and thus, to detect subtle alterations that might indicate early stages or progression of \gls{ad} \cite{Beheshti2022}. However, a diagnosis plan that involves regular \gls{pet} scanning poses challenges, including increased radiation exposure, high cost, and patient burden. Computational models enabling to foresee individual trajectories of brain metabolism could counter this pitfall and thus, significantly impact personalized treatment strategies, early disease management, and therapy control in \gls{ad}. 

Several studies have demonstrated that artificial intelligence can be employed to predict clinical changes of \gls{ad} on the basis of baseline neuroimaging information\cite{Dansson2021, doeringbrainage2024}. However, studies predicting actual longitudinal changes in a specific brain imaging modality are sparse across clinical contexts. \Gls{i2i} methodologies, designed to generate new images from existing ones, hold particular promise for this endeavor, as they make use of the rich information embedded within input images to produce robust image predictions as output. The most popular \gls{i2i} network in biomedical computer vision is the U-Net\cite{Ronneberger2015-ny}, which was designed for medical image segmentation. At its core, U-Net comprises an initial encoding component, where patterns are identified within the input, and a decoding component, allowing to produce patterns for the output. U-Net-like architectures have since been adapted in a variety of medical image generation tasks. Two previous studies employed similar network architectures for the prediction of longitudinal brain changes. Peng et al.\cite{Peng2021} predicted brain changes on MRI in infants using a generative adversarial network, where the generator, i.e., the part of the network that predicted the images, was based on a supervised \gls{cnn}. In the only currently published approach for the longitudinal prediction of \gls{fdg} \gls{pet}, Choi et al.\cite{Choi2018} implemented an unsupervised conditional variational autoencoder, which generated realistic follow-up data in a small sample (\textit{n} = 26) of \gls{cn} subjects based on baseline data and age information. However, the prediction of longitudinal metabolic changes in larger and cognitively heterogeneous samples has not been previously tested. Moreover, supervised learning techniques may enhance the ability to assess individual prediction accuracy in this task. We hypothesize that accurate prediction of \gls{fdg} \gls{pet} in the subsequent year can be achieved by utilizing data from only the two previous years and implementing a time-series prediction paradigm, i.e., training a model to estimate future brain metabolism trajectories. Here, we propose a supervised approach to predict future \gls{fdg} \gls{pet} scans data using a \gls{cnn} equipped with \gls{lstm}.
\section{DATA}
\label{sec:data}
A total of 161 participants (51 \gls{cn}, 95 \gls{mci}, and 15 early dementia) from the \gls{adni} (adni.loni.usc.edu) were included. A diagnosis of \gls{cn} entailed that individuals had no significant impairment in memory, cognitive functions, or activities of daily living. \gls{mci} patients showed measurable impairment in cognitive function in the absence of dementia or significant impairments of daily living. Dementia patients fulfilled the NINCDS/ADRDA criteria for probable AD (https://adni.loni.usc.edu/methods/documents/). To maintain a moderate amount of variance in our sample, we included only individuals who were cognitively unimpaired, or were at most in the early symptomatic stages of dementia (\gls{cdr} score of $\le$ 4)\cite{OBryant2008-wb}. All included individuals received brain \gls{fdg} \gls{pet} scans without image artifacts in three consecutive years. Follow-up \gls{fdg} \gls{pet} data up to seven years after baseline was also obtained whenever available.

\gls{fdg} \gls{pet} scans were downloaded from \gls{adni} with minimal pre-processing (“Co-registered, averaged”-format) and further pre-processing was performed using the Statistical Parametric Mapping 12 toolbox (SPM12 v7771; www.fil.ion.ucl.ac.uk): First, all \gls{fdg} \gls{pet} scans were aligned to the anterior commissure/posterior commissure, and subsequently spatially normalized to standard MNI-152 space, yielding a scan of size 79x95x79 voxels. Second, intensity normalization was performed by calculating the standardized uptake value ratio (SUVR; reference: pons\cite{Verger2021, Nugent2020-pc}) of all \gls{fdg} \gls{pet} scans. Third, all images were brain-masked to prevent background noise from influencing predictive performance, and they were smoothed using a 4x4x4 voxel Gaussian kernel. Finally, to facilitate neural network operations, all \gls{fdg} \gls{pet} scans were padded in each direction to obtain even numbers in each dimension (final size: 80x96x80 voxels).

\section{METHODOLOGY}
\label{sec:method}

\subsection{Neural network}
\label{sec:model}
\begin{figure}[ht]
   \begin{center}
   \frame{\includegraphics[height=8cm]{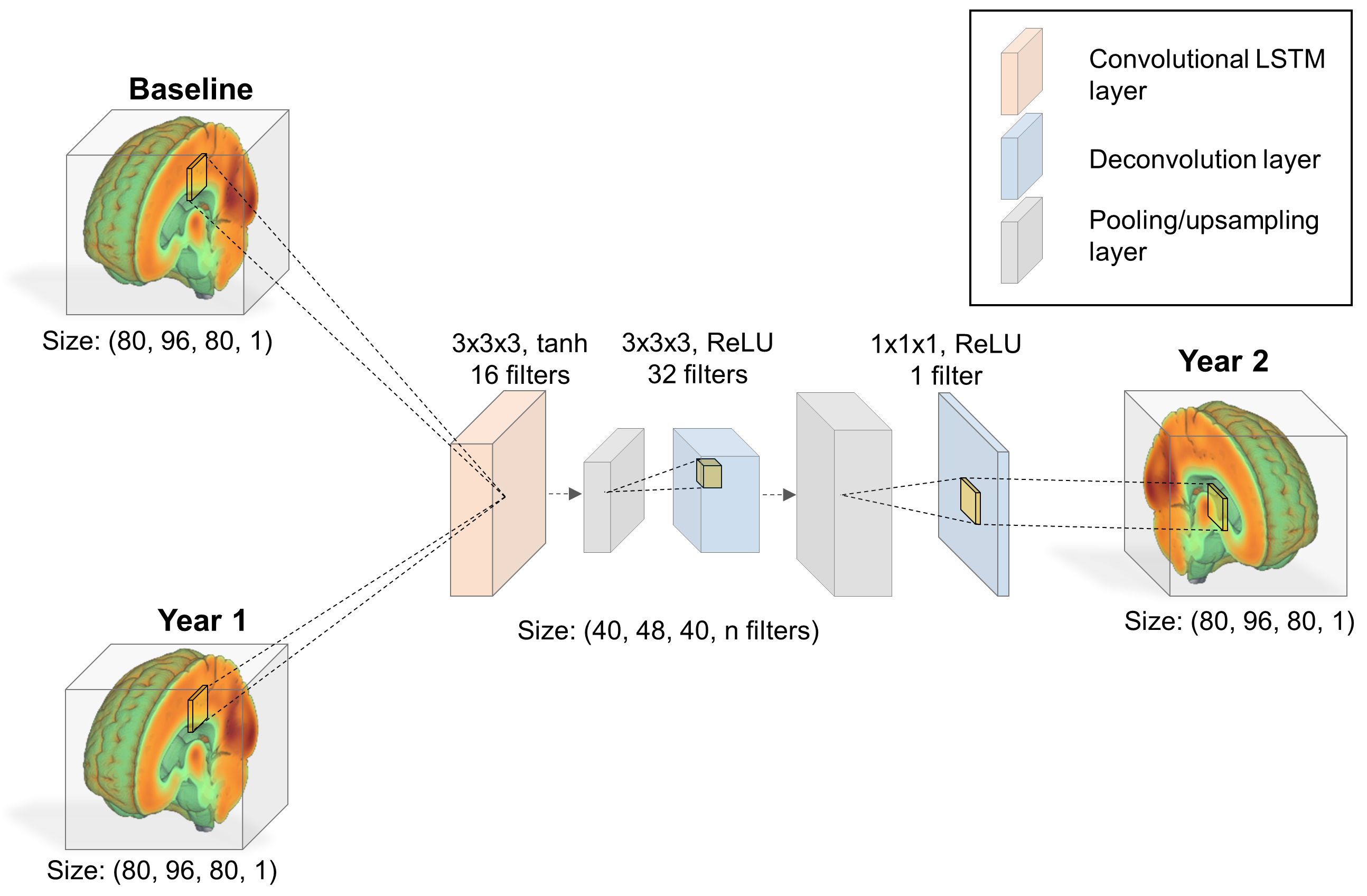}}
   \end{center}
   \caption[model] 
   { \label{fig:model} 
\textbf{The proposed \gls{i2i} network architecture}. Our model takes baseline and year-one \gls{fdg} \gls{pet} data as input and processes them along a three-dimensional CNN-LSTM network architecture, yielding the year-two scan. Size indicates (X, Y, Z, channel/filter dimensions). Network operations are depicted schematically in 2D format for enhanced clarity. I2I = Image-to-image translation; LSTM = Long-short term memory; ReLU = Rectified linear unit; tanh = Hyperbolic tangent.}
   \end{figure}

Our proposed \gls{i2i} network used a \gls{cnn} architecture with a convolutional \gls{lstm} layer to predict future \gls{fdg} \gls{pet} scans from baseline and year one (Fig \ref{fig:model}). The model was implemented in Python (v. 3.6.8) using Keras with TensorFlow (v. 2.6.0) backend. For encoding, a convolutional \gls{lstm} layer with 16 filters was employed as the initial layer of our network to effectively capture both the spatial and temporal dependencies inherent in sequential \gls{fdg} \gls{pet}. Max pooling (pooling window size: 2x2x2 voxels) was utilized to reduce the feature maps to more salient features, and batch normalization was then applied to standardize the feature maps. For decoding, a transposed convolution layer with 32 filters was employed, followed by upsampling and a 1x1x1 convolution in the final layer to efficiently consolidate the feature maps into a single output representation of size 80x96x80. Following common practices in \gls{lstm} and \gls{cnn} research, the \gls{lstm} layer employed a hyperbolic tangent activation function (tanh), and the convolution layers (the hidden and the output layer) used a \gls{relu} activation function. Importantly, we chose \gls{relu} for the output layer, as this activation function yields only positive voxel values and \gls{suvr}s are not expected to be negative. (De-)convolution operations, apart from the last layer, used a kernel size of 3x3x3 voxels. 

\subsection{Training, validation and testing}
\label{sec:training_validation}
The network was trained using five-fold \gls{cv}: In each \gls{cv} fold, 80\% of the data were used to train and validate the model, and 20\% of the data were used to test its generalizability (test sample). Of the training and validation data, in turn, 80\% was used to train the network (training sample) and 20\% were used to validate the model in each epoch (validation sample). To increase our training sample size, we augmented the training data using the Medical Open Network for AI library \cite{MONAI} by applying affine transformations, including random rotation between -$\pi$/18 and $\pi$/18 radians ($\estimates$ -10 to 10 degrees), random zoom by a factor of 0.95 to 1.05 and random shift by up to three voxels into the x, y and/or z direction. Each scan triplet (baseline, year one, and year two) was augmented twice, yielding a final training sample of 309 \gls{fdg} \gls{pet} scans. In each augmentation iteration, the same augmentation was applied to a subject's baseline, year-one and year-two scan to maintain voxel-to-voxel correspondence within-subject. Model training was accomplished using mini-batches of eight scans. The \gls{mae} was used as the loss function, as it is known to be more effective in avoiding the prediction of overly blurry images compared to the mean squared error, which is usually preferred for regression tasks\cite{Zhao2015-ti}. The model was trained for 70 epochs, as preliminary tests suggested that validation performance is relatively stable from this point forward.

\subsection{Linear model}
The prediction of future \gls{fdg} \gls{pet} data, based on information from two distinct time points, could also be approached as a voxel- and patient-specific linear regression analysis. To test whether the application of complex neural networks is necessary for the task at hand, we implemented a linear model for comparison. In this approach, the same differences observed between baseline and year one were expected to occur between year one and year two (and subsequent years for longitudinal predictions), by computing for each voxel:
\begin{equation}
    \text{Year}_{X} = \text{Year}_{X-1} - (\text{Year}_{X-2} - \text{Year}_{X-1})
\end{equation}

\subsection{Longitudinal prediction}
Finally, we investigated whether our models could be used to predict brain metabolism beyond year two. Towards this end, we recursively used predicted scans of the two previous years to predict the scan of the subsequent year. For example, to predict the scan in year three, we used the year-one scan and the predicted year two scan as input to our \gls{i2i} network or linear model. Importantly, the cross-validation procedure of the \gls{i2i} network training yielded five different models. To ensure accurate longitudinal predictions and prevent feature leakage, we utilized the model in which each respective individual was part of the test set to predict longitudinal scans. Longitudinal predictions for which there were no corresponding ground truth scans were not evaluated.

\subsection{Systematic evaluation of model performance}
\label{sec:meth_systematic}
The proposed \gls{i2i} network was evaluated on independent test and longitudinal data, focusing on two key aspects: 
\begin{enumerate}
    \item To evaluate the \textbf{overall prediction performance} of our model, we computed the MAE (range: 0 (best) - $\infty$ (worst)) and the structural similarity index (SSIM, range: 0 (worst) - 1 (best))\cite{Wang2004-wq} between each predicted and ground truth scan in years 2-7. MAE and SSIM were then compared across scans predicted by the \gls{i2i} network and the linear model using the non-parametric Wilcoxon signed-rank test, as the performance metrics were not normally distributed across years. Moreover, to obtain an understanding of the spatial distribution of prediction errors, i.e., to test whether \gls{i2i} network prediction errors were systematically, spatially constrained to specific brain regions, we quantified the MAE in different brain regions using the \gls{aal} atlas for segmentation \cite{Tzourio-Mazoyer2002}. 
    \item To evaluate the \textbf{group-specific prediction performance in an AD-specific \gls{roi}}, we computed the mean \gls{suvr} value in a meta-\gls{roi}, known to be prone to hypometabolism in \gls{ad}\cite{Jack2016}, comprising the angular gyrus, posterior cingulate, and inferior temporal cortical cortex. Mean \gls{suvr}s were computed based on ground truth \gls{fdg} \gls{pet} scans, or scans predicted by the \gls{i2i} network or linear model, and subsequently compared using a mixed-model \gls{anova} paradigm. In the \gls{anova}, we used model type (ground truth, \gls{i2i} network, or linear model) as the within-subject, and the diagnostic group as the between-subject factor. The \gls{anova} tests were followed by post-hoc paired t-tests between ground truth scans predicted by either prediction model. The post-hoc tests were evaluated with a Bonferroni-corrected $\alpha$ (year 2: $\alpha$=0.05/6=0.0083; subsequent years (no dementia sample): $\alpha$=0.05/4=0.0125).
    
\end{enumerate}

\section{RESULTS}
\label{sec:results}
\begin{table}[h]
\centering
\caption{Participant characteristics.}
\label{tab:table1}
\begin{tabular}{l c c c r}
\hline
 & \textbf{CN} & \textbf{MCI} & \textbf{Dementia} & \textit{P} value \\ \hline
 \textit{n} in year 0 - 2 & 51 & 95 & 15 & - \\ 
 \textit{n} in year 3 & 41 & 61 & 0 & - \\
 \textit{n} in year 4 & 15 & 19 & 0 & - \\
 \textit{n} in year 5 & 10 & 12 & 0 & - \\
 \textit{n} in year 6 & 8 & 12 & 0 & - \\
 \textit{n} in year 7 & 10 & 10 & 0 & - \\
Age [Y, mean (SD)] & 75.9 (4.2) & 75.5 (7.4) & 76.7 (6.4) & .25 \\ 
Sex [\% female] & 28 & 35 & 33 & .39 \\ 
MMSE [mean (SD)] & 29 (1.1) & 27 (1.7) & 24 (1.6) & $<$ .001 \\
CDR-SOB & 0.0 (0.1) & 1.6 (0.7) & 2.9 (1.1) & $<$ .001 \\
CSF A$\beta_{1-42}$ & 1237.5 (696.2) & 849.4 (484.0) & 877.6 (730.7) & $<$ .001 \\
Years of education [mean (SD)] & 16 (3.1) & 16 (2.8) & 14 (3.7) & .03 \\
\hline
\multicolumn{5}{p{14.5cm}}{\scriptsize\textit{Notes.} Number of participants is reported as the number of participants for our main analyses (baseline (year 0), year 1 and year 2) and for subsequent years. Participants were not required to have follow-up data for each longitudinal year. Age, cognitive scores (MMSE and CDR-SOB) and $\beta$-amyloid levels in cerebrospinal fluid (CSF) are reported from baseline. \textit{P} values obtained using a one-way \gls{anova} (numerical) or pair-wise $\chi^2$ tests (categorical). Smallest p-value is indicated for multiple categorical comparisons. MMSE = Mini Mental State Examination; SD = Standard deviation.}
\end{tabular}
\vspace{0.2em}
\end{table}

\begin{figure}[h]
   \begin{center}
   \frame{\includegraphics[height=10cm]{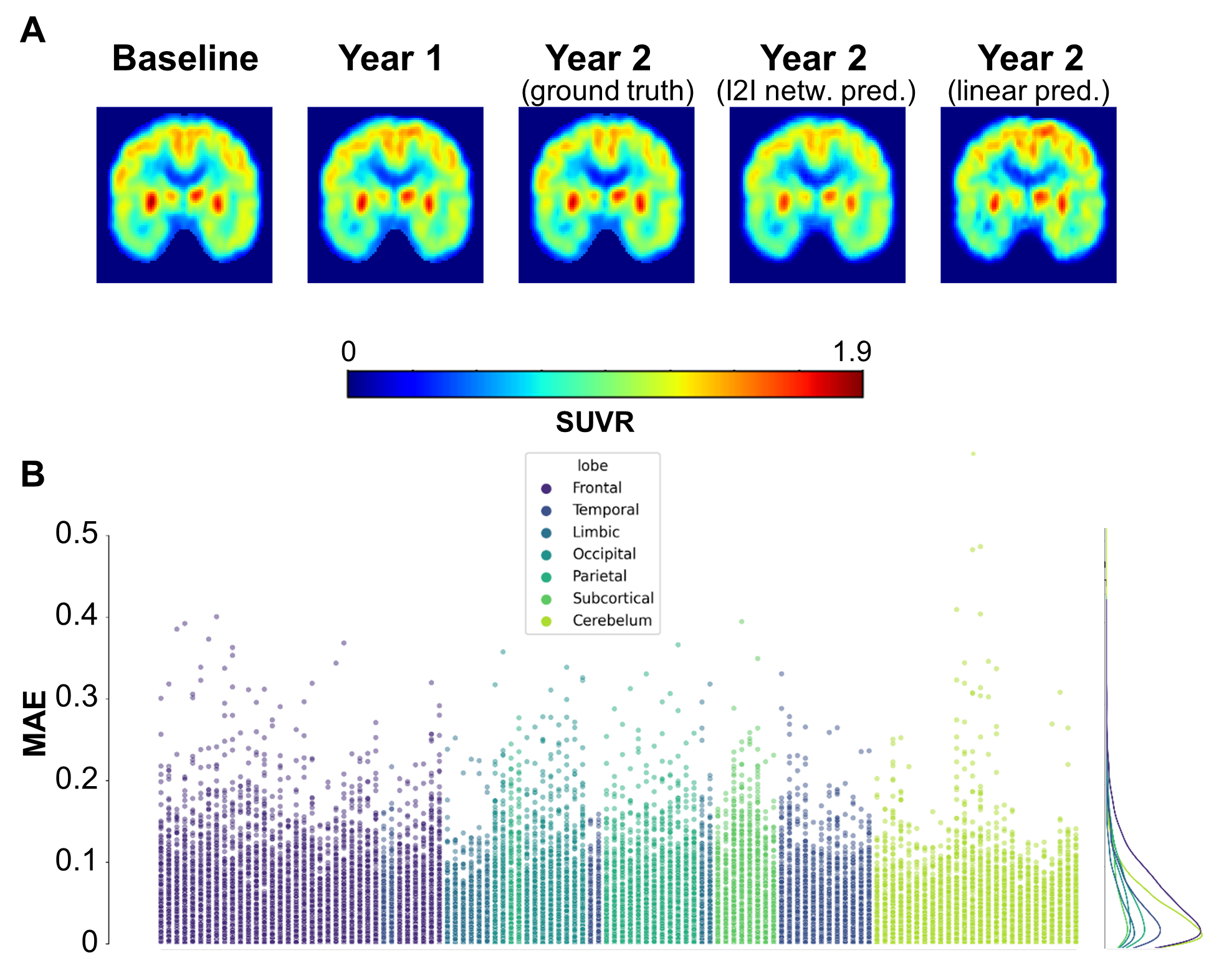}}
   \end{center}
   \caption[results] 
   { \label{fig:accuracy} 
\textbf{Results from our \gls{i2i} network and the linear model.} A: Example baseline, year-one, year-two and predicted (pred.) scans of an MCI patient. B: Visualization of MAE in brain regions, colored by lobe. Each column represents one \gls{aal} atlas region. The density plots on the left summarize brain regions by lobe, indicating that MAE was low across lobes.}
\vspace{-1.5em}
   \end{figure}

\subsection{Participants}
This study included 161 participants. Table \ref{tab:table1} presents an overview of participant characteristics. The diagnostic groups did not significantly differ with respect to age or sex, however, dementia patients had fewer years of education compared to \gls{cn} and \gls{mci} subjects. By design, all three groups significantly differed with respect to baseline cognitive performance. Dementia patients only had \gls{fdg} \gls{pet} data from three subsequent years, i.e., no longitudinal analyses could be conducted.

\subsection{Overall prediction performance}

\begin{figure}[b]
   \begin{center}
   \frame{\includegraphics[height=11.5cm]{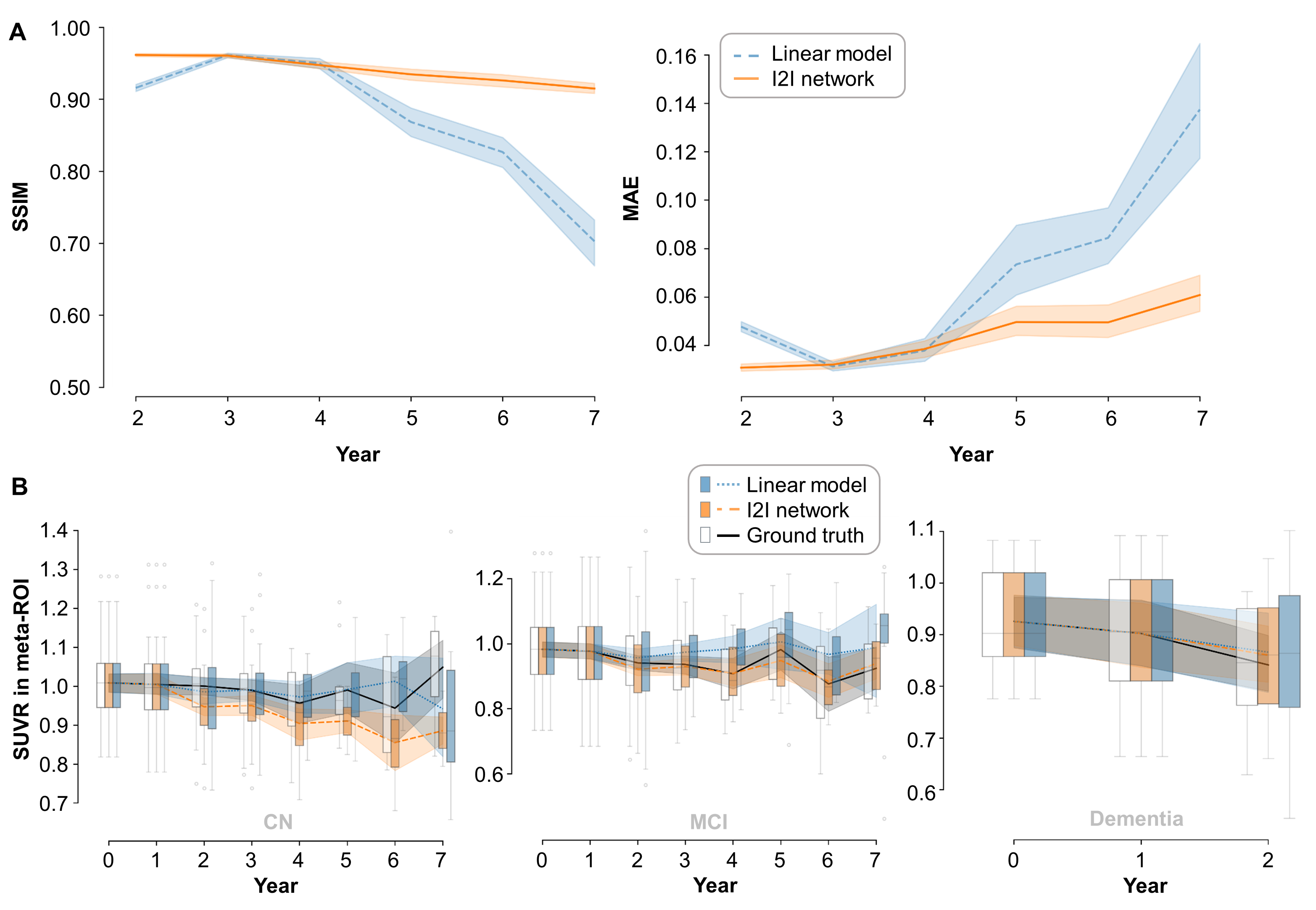}}
   \end{center}
   \caption[results] 
   { \label{fig:longitudinal} 
\textbf{Model performance across years.} A: Prediction performance  of the \gls{i2i} network (orange) was stable across years, while predictions of the linear model (blue) were only accurate in year three and four. B: Predictions yielded by the \gls{i2i} network (orange) were highly similar to ground truth (white boxes, gray lines) in MCI patients. In dementia patients, predictions were comparably accurate using either model (data only available until year 2), while the linear model alone (blue) was superior in predicting future scans in CN. CN = cognitively normal; I2I = Image-to-image translation; MAE = mean absolute error; MCI = mild cognitive impairment; SSIM = structural similarity index; ROI = region of interest; SUVR = standardized uptake value ratio.}
   \end{figure}
Our \gls{i2i} network achieved high accuracy in the prediction of year two, as evidenced in a small voxel-wise error (MAE = 0.031) and high structural similarity (SSIM = 0.961). In comparison, the linear model yielded higher errors (MAE = 0.048, SSIM = 0.916; Fig. \ref{fig:accuracy} and \ref{fig:longitudinal}A). By recursively utilizing the network's predictions as inputs for the prediction of subsequent years, our model achieved highly accurate predictions of longitudinal \gls{fdg} \gls{pet} scans (year 3: MAE = 0.032, SSIM = 0.960; year 4: MAE = 0.038, SSIM = 0.947; year 5: MAE = 0.050, SSIM = 0.934; year 6: MAE = 0.049, SSIM = 0.926; year 7: MAE = 0.061, SSIM = 0.915). The linear model performed well in year 3 (MAE = 0.032, SSIM = 0.961) and 4 (MAE = 0.038, SSIM = 0.950), but showed a steep increase in prediction error thereafter (year 5: MAE = 0.073, SSIM = 0.868; year 6: MAE = 0.084, SSIM = 0.827; year 7: MAE = 0.137, SSIM = 0.703; Fig. \ref{fig:longitudinal}A). Scans predicted by the \gls{i2i} network were more similar to ground truth scans in most years with respect to MAE (year 2, 5, 6 and 7: W=0, \textit{P}$<$.001) and SSIM (year 2: W=256, \textit{P}$<$.001; year 5: W=35, \textit{P}=.002; year 6: W=6, \textit{P}$<$.001, year 7: W=0, \textit{P}$<$.001), while the SSIM (but not MAE) of linear-model predicted scans was slightly lower compared to the \gls{i2i} network in year 3 (MAE: W=2337, \textit{P}=.335; SSIM: W=2001, \textit{P}=.037) and no significant difference was detected in year 4 (MAE: W=258, \textit{P}=.505; SSIM: W=221, \textit{P}=.194). Region-wise analysis of prediction errors yielded that there was no systematic, spatial bias towards high MAE in specific brain regions, as the MAE was low across all atlas regions (Fig. \ref{fig:accuracy}B).

\subsection{Group-specific prediction performance in an AD-specific region-of-interest}
\begin{table}[b]
\centering

\begin{tabular}{llcccc}
 \multicolumn{6}{p{10,8cm}}{\caption{Comparison of ground truth and predicted \gls{suvr}s in meta-ROI across groups and models (mixed-model ANOVA) and within-groups across models (post-hoc paired t-test).\label{tab:anovas}}}\\

\hline
\textbf{Year} & \textbf{Comparison} & \multicolumn{2}{c}{\textbf{GT-I2I netw.}} & \multicolumn{2}{c}{\textbf{GT-linear}} \\
\hline
\textbf{Y2} & ANOVA & \multicolumn{4}{c}{Group: $\star$; Model: $\star\star$; Group*Model: $\star\star$} \\
 & \hspace{.3cm}CN & t=6.73 & \textit{P}$<$0.001 & t=1.33 & \textit{P}=0.191 \\
 & \hspace{.3cm}MCI & t=3.98 & \textit{P}$<$0.001 & t=-2.03 & \textit{P}=0.045 \\  \vspace{.3em}
 & \hspace{.3cm}Dementia & t=1.93 & \textit{P}=0.075 & t=-1.45 & \textit{P}=0.170 \\
\textbf{Y3} & ANOVA & \multicolumn{4}{c}{Group: n.s.; Model: $\star\star$; Group*Model: $\star\star$} \\
 & \hspace{.3cm}CN & t=4.94 & \textit{P}$<$0.001 & t=-0.08 & \textit{P}=0.935 \\ \vspace{.3em}
 & \hspace{.3cm}MCI & t=0.83 & \textit{P}=0.408 & t=-5.05 & \textit{P}$<$0.001 \\
 \textbf{Y4} & ANOVA & \multicolumn{4}{c}{Group: n.s.; Model: $\star\star$; Group*Model: $\star$} \\
 & \hspace{.3cm}CN & t=3.56 & \textit{P}=0.003 & t=-1.32 & \textit{P}=0.201 \\ \vspace{.3em}
 & \hspace{.3cm}MCI & t=-0.08 & \textit{P}=0.938 & t=-5.05 & \textit{P}=0.003 \\
\textbf{Y5} & ANOVA & \multicolumn{4}{c}{Group: n.s.; Model: $\star$; Group*Model: n.s.} \\
 & \hspace{.3cm}CN & t=2.74 & \textit{P}=0.023 & t=-0.03 & \textit{P}=0.977 \\ \vspace{.3em}
 & \hspace{.3cm}MCI & t=1.12 & \textit{P}=0.289 & t=-0.543 & \textit{P}=0.598 \\
\textbf{Y6} & ANCOVA & \multicolumn{4}{c}{Group: n.s.; Model: $\star$; Group*Model: n.s.} \\
 & \hspace{.3cm}CN & t=2.42 & \textit{P}=0.046 & t=-1.84 & \textit{P}=0.109 \\ \vspace{.3em}
 & \hspace{.3cm}MCI & t=-0.24 & \textit{P}=0.812 & t=-1.32 & \textit{P}=0.212 \\
\textbf{Y7} & ANOVA & \multicolumn{4}{c}{Group: n.s.; Model: n.s.; Group*Model: n.s.} \\
 & \hspace{.3cm}CN & t=5.98 & \textit{P}$<$0.001 & t=1.45 & \textit{P}=0.182 \\
 & \hspace{.3cm}MCI & t=-0.46 & \textit{P}=0.657 & t=-0.78 & \textit{P}=0.457 \\
 \hline
\multicolumn{6}{p{10,8cm}}{\scriptsize\textit{Notes.} Comparisons derived using AD-susceptible meta-ROI described in section \ref{sec:meth_systematic}. Group*Model represents the interaction term between group and model. ANOVA significance codes: $\star$ \textit{P}$<$0.05, $\star\star$ \textit{P}$<$0.01., n.s. not significant} \\

\end{tabular}
\end{table}
Next, we computed the mean \gls{suvr} signal in an \gls{ad}-susceptible meta-\gls{roi} from ground truth scans, as well as from scans predicted by the \gls{i2i} network or the linear model. A mixed-model ANOVA was used to compare meta-\gls{roi} \gls{suvr}s within-groups, across models (with ``ground truth'' also considered as a ``model'' for simplicity). Table \ref{tab:anovas} and Figure \ref{fig:longitudinal}B show that meta-\gls{roi} \gls{suvr}s from scans predicted by our \gls{i2i} network were highly accurate in MCI across years, while the linear model yielded more accurate estimates of meta-\gls{roi} \gls{suvr} in CN individuals. Both prediction models predicted close-to-ground truth meta-\gls{roi} \gls{suvr}s in the dementia sample. 

\label{sec:clinsig}
\section{DISCUSSION}
\label{sec:conclusion}
In this study, we developed and validated an \gls{i2i} network, composed of \gls{cnn} and \gls{lstm} layers, for the prediction of future \gls{fdg} \gls{pet} scans using baseline and year-one data. Notably, the network demonstrated high predictive accuracy not only for year two but extending up to year seven post-baseline. Moreover, predictions were significantly more accurate compared to a simpler, linear model in most years. While our model predicted future brain metabolism in AD-susceptible regions well in MCI and dementia patients, it was outperformed in this task by the linear model in CN.

Surprisingly, the linear model, which applied the same voxel-wise differences observed in previous years to predict the \gls{fdg} \gls{pet} scan for the next, was sufficient to predict development of \gls{fdg} \gls{pet} in \gls{cn} individuals. This observation suggests that metabolic changes over time in \gls{cn} subjects are very subtle \cite{Landau2012-na} and linear. For clinical practice, this finding presents a simple tool for the assessment of future brain metabolic decline in CN, and it provides evidence that missing longitudinal data of CN may be imputed based simply on voxel-wise linear predictions. However, due to the relatively stable brain metabolism in our \gls{cn} sample, the success of the linear model may not generalize to other objectively unimpaired populations or conditions with more dynamic metabolic changes (e.g., individuals with subjective cognitive decline). Our proposed \gls{i2i} network was not able to predict the absence of profound changes observed in \gls{cn} over time and predicted a comparably steep decline in brain metabolism. This may have resulted from data imbalance, as most individuals in our sample were \gls{mci} patients, where some degree of neurodegeneration is expected\cite{Caminiti2023}. In fact, the \gls{i2i} network was highly accurate in predicting future \gls{fdg} \gls{pet} in \gls{mci} patients, with no significant difference between ground truth and predicted scans from year three forward. Importantly, the prediction of future \gls{fdg} \gls{pet} was based on only two time points which may be located at very different positions on the overall trajectory in individual subjects. In healthy subjects, \gls{fdg} \gls{pet} scans are likely nearly the same between baseline and subsequent years, independent of when they are acquired. In patients with beginning neurodegeneration, i.e., \gls{mci} patients, on the other hand, the magnitude of neurodegeneration observed between any two time points may be modified by disease duration and subtype of \gls{mci}\cite{Caminiti2023}. Thus, changes may be non-linear both across and within individuals over time. In contrast to the linear model, the advantage of neural networks lies in their ability to detect and model such complex, nonlinear interactions within the data, which appeared to enable the prediction of brain metabolic decline in \gls{mci} for up to seven years after the initial visit. For dementia patients, both the linear model and the \gls{i2i} network accurately predicted brain metabolic decline in year two, suggesting brain metabolic changes in advanced neurodegeneration may also follow a close-to-linear trajectory. While both suggested models are theoretically capable of modeling linear changes, the less complex and patient-specific linear model may be preferable and sufficient to assess brain metabolic decline in dementia patients in clinical practice. 

Our results yield several clinical implications. To clinicians, these models could offer valuable prognostic insights, guiding early intervention strategies to preemptively address anticipated declines in brain metabolism. For patients, an approach like the current may facilitate diagnostic procedures by alleviating the need for frequent monitoring. In the realm of clinical trials, predictive modeling as proposed here could be useful by identifying patients most likely to show a steep trajectory of neurodegeneration in the future, and who, in turn, may benefit most from a specific medication. Furthermore, the model's ability to generate personalized predictions of metabolic decline could serve as a benchmark for evaluating the efficacy of new drugs, comparing observed post-treatment brain metabolism against predicted outcomes.

Some limitations of our study must be acknowledged. First, the dataset used for training and evaluation was small compared to common deep learning datasets, and it was also limited in diversity\cite{Ashford2022}. Moreover, our model did not consider demographic and clinical information, which may be vital for predicting the development of \gls{fdg} \gls{pet} with even higher precision. These variables were not included to provide an unbiased proof-of-principle account on the potential of an \gls{i2i} network and patient-specific linear models to predict future brain metabolic decline. Encoding demographic or clinical information within the image, or introducing them in a linear latent space, e.g., mimicking the architecture of an autoencoder, may allow future studies to assess the potential of predicting brain metabolic decline under consideration of such information\cite{Choi2018}. Moreover, across subjects, in addition to regions showing metabolic decline, we also noted some longitudinal increase of brain metabolism in individual other regions in the ground truth data. This may reflect true, e.g. compensatory effects but it cannot be excluded that these findings are conditioned by the choice of the reference region. We chose the pons as the reference region based on its established value and high sensitivity to brain metabolic changes in \gls{ad}\cite{Nugent2020-pc} and aging\cite{Verger2021}, demonstrated in previous studies. However, these studies predominantly were of cross-sectional nature and longitudinal changes affecting the pons itself cannot be excluded. Possibly, other reference regions may be more suitable to capture longitudinal changes, although this would also apply to repeated longitudinal PET measurements accordingly. To investigate the optimal reference region to monitor brain metabolism over time thus constitutes an important direction for future studies. Furthermore, we only assessed the most common meta-ROI for \gls{fdg} \gls{pet} and evaluation of other \gls{roi}s may further elucidate on the potential of either model to predict future brain metabolism changes. Finally, we used a relatively shallow architecture, which was able to maintain high MAE and structural similarity until year seven. However, this architecture may have not been able to grasp the full complexity of brain metabolic changes over time, and consequently have prevented stronger predictive performance in CN. The architecture proposed here is one of many possible, and future studies could systematically investigate whether more complex architectures better capture brain metabolism trajectories across clinical groups. 

In conclusion, this proof-of-principle study highlighted the potential of convolutional \gls{i2i} networks and linear models for the assessment of future brain metabolic decline years in advance. Once further validated, the proposed models might contribute to the enhancement of clinical management of neurodegenerative disorders and streamlining of the design and execution of clinical trials.

\section*{ACKNOWLEDGEMENTS}
Data collection and sharing for this project was funded by the ADNI (National Institutes of Health Grant U01 AG024904) and Department of Defense ADNI. This work was partially funded by the German Academic Exchange Service (DAAD). We furthermore thank the Regional Computing Center of the University of Cologne (RRZK) for providing computing time on the DFG-funded (Funding number: INST 216/512/1FUGG) High-Performance Computing (HPC) system CHEOPS as well as support. Moreover, we are grateful for the generous travel grant by the DZNE Foundation to attend SPIE Medical Imaging.

\hspace{-0.4cm}* Data used in preparation of this article were obtained from the Alzheimer’s Disease Neuroimaging Initiative (ADNI) database (adni.loni.usc.edu). As such, the investigators within the ADNI contributed to the design and implementation of ADNI and/or provided data but did not participate in analysis or writing of this report. A complete listing of ADNI investigators can be found at: \url{http://adni.loni.usc.edu/wp-content/uploads/how_to_apply/ADNI_Acknowledgement_List.pdf}

\bibliography{report} 

\begin{thebibliography}{10}

\bibitem{Caminiti2023}
Caminiti, S.~P., De~Francesco, S., Tondo, G., Galli, A., Redolfi, A., Perani, D., {Alzheimer's Disease Neuroimaging Initiative}, and {Interceptor Project}, ``{FDG-PET markers of heterogeneity and different risk of progression in amnestic MCI},'' {\em Alzheimer's and dementia}  (2023).

\bibitem{Mosconi2010}
Mosconi, L., Berti, V., Glodzik, L., Pupi, A., {De Santi}, S., and {De Leon}, M.~J., ``{Pre-clinical detection of Alzheimer's disease using FDG-PET, with or without amyloid imaging},'' (2010).

\bibitem{Nestor2018}
Nestor, P.~J., Altomare, D., Festari, C., Drzezga, A., Rivolta, J., Walker, Z., Bouwman, F., Orini, S., Law, I., Agosta, F., Arbizu, J., Boccardi, M., Nobili, F., and Frisoni, G.~B., ``{Clinical utility of FDG-PET for the differential diagnosis among the main forms of dementia},'' {\em European Journal of Nuclear Medicine and Molecular Imaging}~{\bf 45}(9) (2018).

\bibitem{Beheshti2022}
Beheshti, I., Geddert, N., Perron, J., Gupta, V., Albensi, B.~C., and Ko, J.~H., ``{Monitoring Alzheimer's Disease Progression in Mild Cognitive Impairment Stage Using Machine Learning-Based FDG-PET Classification Methods},'' {\em Journal of Alzheimer's Disease}~{\bf 89}(4) (2022).

\bibitem{Dansson2021}
Dansson, H.~V., Stempfle, L., Egilsd{\'{o}}ttir, H., Schliep, A., Portelius, E., Blennow, K., Zetterberg, H., and Johansson, F.~D., ``{Predicting progression and cognitive decline in amyloid-positive patients with Alzheimer's disease},'' {\em Alzheimer's Research and Therapy}~{\bf 13}(1) (2021).

\bibitem{doeringbrainage2024}
Doering, E., Antonopoulos, G., Hoenig, M., van Eimeren, T., Daamen, M., Boecker, H., Jessen, F., D{\"u}zel, E., Eickhoff, S., Patil, K., and Drzezga, A., ``{MRI} or $^{18}${F-FDG PET} for brain age gap estimation: Links to cognition, pathology, and {A}lzheimer disease progression,'' {\em Journal of Nuclear Medicine}  (2023).

\bibitem{Ronneberger2015-ny}
Ronneberger, O., Fischer, P., and Brox, T., ``{U-Net}: Convolutional networks for biomedical image segmentation,'' (2015).

\bibitem{Peng2021}
Peng, L., Lin, L., Lin, Y., Chen, Y.~W., Mo, Z., Vlasova, R.~M., Kim, S.~H., Evans, A.~C., Dager, S.~R., Estes, A.~M., McKinstry, R.~C., Botteron, K.~N., Gerig, G., Schultz, R.~T., Hazlett, H.~C., Piven, J., Burrows, C.~A., Grzadzinski, R.~L., Girault, J.~B., Shen, M.~D., and Styner, M.~A., ``{Longitudinal Prediction of Infant MR Images With Multi-Contrast Perceptual Adversarial Learning},'' {\em Frontiers in Neuroscience}~{\bf 15} (2021).

\bibitem{Choi2018}
Choi, H., Kang, H., and Lee, D.~S., ``{Predicting aging of brain metabolic topography using variational autoencoder},'' {\em Frontiers in Aging Neuroscience}~{\bf 10}(JUL) (2018).

\bibitem{OBryant2008-wb}
O'Bryant, S.~E., Waring, S.~C., Cullum, C.~M., Hall, J., Lacritz, L., Massman, P.~J., Lupo, P.~J., Reisch, J.~S., Doody, R., and {Texas Alzheimer's Research Consortium}, ``{Staging dementia using Clinical Dementia Rating Scale Sum of Boxes scores: a Texas Alzheimer's research consortium study},'' {\em Arch. Neurol.}~{\bf 65}(8),  1091--1095 (2008).

\bibitem{Verger2021}
Verger, A., Doyen, M., Campion, J.~Y., and Guedj, E., ``{The pons as reference region for intensity normalization in semi-quantitative analysis of brain 18FDG PET: application to metabolic changes related to ageing in conventional and digital control databases},'' {\em EJNMMI Research}~{\bf 11}(1),  1--7 (2021).

\bibitem{Nugent2020-pc}
Nugent, S., Croteau, E., Potvin, O., Castellano, C.-A., Dieumegarde, L., Cunnane, S.~C., and Duchesne, S., ``Selection of the optimal intensity normalization region for {FDG-PET} studies of normal aging and alzheimer's disease,'' {\em Sci. Rep.}~{\bf 10},  9261 (June 2020).

\bibitem{MONAI}
{The MONAI Consortium}, ``Monai: Medical open network for ai.'' \url{https://github.com/Project-MONAI/MONAI} (2020).

\bibitem{Zhao2015-ti}
Zhao, H., Gallo, O., Frosio, I., and Kautz, J., ``Loss functions for neural networks for image processing,'' (2015).

\bibitem{Wang2004-wq}
Wang, Z., Bovik, A.~C., Sheikh, H.~R., and Simoncelli, E.~P., ``Image quality assessment: from error visibility to structural similarity,'' {\em IEEE Trans. Image Process.}~{\bf 13},  600--612 (Apr. 2004).

\bibitem{Tzourio-Mazoyer2002}
Tzourio-Mazoyer, N., Landeau, B., Papathanassiou, D., Crivello, F., Etard, O., Delcroix, N., Mazoyer, B., and Joliot, M., ``{Automated anatomical labeling of activations in SPM using a macroscopic anatomical parcellation of the MNI MRI single-subject brain},'' {\em NeuroImage}~{\bf 15}(1) (2002).

\bibitem{Jack2016}
Jack, C.~R., Wiste, H.~J., Weigand, S.~D., Therneau, T.~M., Lowe, V.~J., Knopman, D.~S., Gunter, J.~L., Senjem, M.~L., Jones, D.~T., Kantarci, K., Machulda, M.~M., Mielke, M.~M., Roberts, R.~O., Vemuri, P., Reyes, D.~A., and Petersen, R.~C., ``Defining imaging biomarker cut points for brain aging and alzheimer’s disease,'' {\em Alzheimer’s and Dementia}~{\bf 13}(3),  205–216 (2016).

\bibitem{Landau2012-na}
Landau, S.~M., Mintun, M.~A., Joshi, A.~D., Koeppe, R.~A., Petersen, R.~C., Aisen, P.~S., Weiner, M.~W., Jagust, W.~J., and {Alzheimer's Disease Neuroimaging Initiative}, ``Amyloid deposition, hypometabolism, and longitudinal cognitive decline,'' {\em Ann. Neurol.}~{\bf 72},  578--586 (Oct. 2012).

\bibitem{Ashford2022}
Ashford, M.~T., Raman, R., Miller, G., Donohue, M.~C., Okonkwo, O.~C., Mindt, M.~R., Nosheny, R.~L., Coker, G.~A., Petersen, R.~C., Aisen, P.~S., and Weiner, M.~W., ``{Screening and enrollment of underrepresented ethnocultural and educational populations in the Alzheimer's Disease Neuroimaging Initiative (ADNI)},'' {\em Alzheimer's and Dementia}~{\bf 18}(12) (2022).

\end{thebibliography}
\bibliographystyle{spiebib} 

\end{document}